\documentclass[%
 reprint,
 amsmath,amssymb,
 aps, prl
]{revtex4-2}
\usepackage{graphicx}
\usepackage{textcomp}
\usepackage{amssymb}
\usepackage[colorlinks,linkcolor=blue,anchorcolor=blue,citecolor=blue,urlcolor=black]{hyperref}
\usepackage[utf8]{inputenc}
\usepackage{amsmath}
\usepackage{tipa}
\usepackage{comment}
\usepackage{tabularray}
\usepackage{mathtools}
\usepackage{braket}
\usepackage{listings}
\usepackage{xcolor}
\usepackage{soul}
\usepackage{etoolbox}
\usepackage{lineno}
\usepackage{ulem}
\usepackage{svg}
\usepackage{CJKutf8}
\usepackage{multirow}
\usepackage{tabularray}
\begin{document}

\title{Comment on ``Possibility of superradiant neutrino emission by atomic condensate'' by M. Blasone, L. Gastaldo and F. Romeo, Phys. Rev. D 113, 053010 (2026)}

\author{Wolfgang Ketterle}
\affiliation{Department of Physics, Massachusetts Institute of Technology, Cambridge, MA 02139, USA}
\affiliation{Research Laboratory of Electronics, Massachusetts Institute of Technology, Cambridge, MA 02139, USA}
\affiliation{MIT-Harvard Center for Ultracold Atoms, Cambridge, MA, USA}

\author{Hanzhen Lin }
\affiliation{Department of Physics, Massachusetts Institute of Technology, Cambridge, MA 02139, USA}
\affiliation{Research Laboratory of Electronics, Massachusetts Institute of Technology, Cambridge, MA 02139, USA}
\affiliation{MIT-Harvard Center for Ultracold Atoms, Cambridge, MA, USA}

\author{Yu-Kun Lu}
\affiliation{Department of Physics, Massachusetts Institute of Technology, Cambridge, MA 02139, USA}
\affiliation{Research Laboratory of Electronics, Massachusetts Institute of Technology, Cambridge, MA 02139, USA}
\affiliation{MIT-Harvard Center for Ultracold Atoms, Cambridge, MA, USA}

\begin{abstract}
We show that the recent proposal for superradiant emission of neutrinos~\cite{neutrino_laser_blasone_etal} cannot evade our proof~\cite{fermion_impossibility} that superradiant neutrino emission is fundamentally impossible. Pairing two fermions in a molecule does not remove the cancellation of interference terms in neutrino emission due to fermionic anticommutators.

\end{abstract}
\maketitle
%%%%%%%%%%%%%%%%%%%%%%%%%%%%%%%%

This paper describes possibilities to achieve superradiant neutrino emission from atomic Bose-Einstein condensates.   All these possibilities are rigorously ruled out by proofs and theorems in our papers~\cite{fermion_impossibility,boson_impossibility}. None of the three models presented in this paper correctly describes neutrino emission. 
More specifically, we disagree on the following four points:

\textit{(1)} As summarized in~\cite{neutrino_laser_blasone_etal} neutrino emission is caused by electron capture, leading to an atom with inner shell excitation which decays by X ray emission within 1 femtosecond. Furthermore, the daughter atom has recoil due to the neutrino emission and populates $\approx 10^{12}$ modes (for 1 MeV neutrinos), and $\approx 10^6$ modes for 1 keV neutrinos (which can be created by Ho-163 decay \cite{neutrino_laser_blasone_etal}). In~\cite{boson_impossibility} we show that these two effects lower any possible gain by a factor of $\approx 10^{17}$ (for 1 MeV neutrinos) or $\approx 10^8$ for keV neutrinos).

Therefore, the kinetics of the emission of keV or MeV neutrinos eliminates the possibility of any superradiant enhancement of neutrino emission in a quantum gas. Any form of superradiance (including all scenarios discussed in~\cite{neutrino_laser_blasone_etal,neutrino_laser_jones_formaggio}) relies on stimulation via coherence between initial and final states. For electron capture processes, the escape and/or further rapid decay of the reaction products and the competition between many modes prevents the build-up of any coherence.

The discussion in  Ref.~\cite{neutrino_laser_blasone_etal} therefore addresses only a hypothetical question: what forms of neutrino superradiance are possible by symmetry, assuming neutrino emission at sub-eV energies (to allow approximations with only one or a few modes) and ignoring the short lifetime of the final state of an electron capture process. None of the atomic isotopes discussed by Blasone et al. are possible candidates.

\textit{(2)} The first part of the paper discusses the possibility of ``Neutrino superradiance in a bosonic-bosonic model''~\cite{neutrino_laser_blasone_etal}.  Here the authors discuss neutrino emission between two bosonic states and conclude that neutrino superradiance is possible. However, two bosonic states cannot be connected by neutrino emission, only by emission of a photon or gamma ray.  Indeed, what the authors derive are the standard equations for photonic superradiance (Eq. 12 and 15). Their conclusion that ``this idealized situation is not realized in practice'' is misleading:  their model is in violation of fundamental physics laws and has nothing to do with neutrino emission.

\textit{(3)} Their chapter on neutrino emission in a bosonic-fermionic model is consistent with fundamental symmetries, but it does not correctly describe neutrino emission. In Eq. (20) they incorrectly use a jump operator from a bosonic state to two fermionic states, in contrast to their Eq. (1) which describes the reaction correctly by a Hamiltonian term. A Lindblad jump operator describes the physics of a reservoir (to accept energy or particles). Their Eq. (19) represents the Hamiltonian for their system which consists of bosonic parent atoms, fermionic daughter atoms, and the neutrinos. Therefore, the neutrino emission must be described by a unitary interaction term within the system, and not a jump operator.

%\sout{ Since they assume the reaction or jump can only populate a single fermionic final state, they obtain the standard Pauli blocking equations. This does not reproduce the result in our paper~\cite{fermion_impossibility}, where we have shown that neutrino superradiance is impossible even in the absence of Pauli blocking by the fermionic daughter atom.}

\textit{(4)} Most importantly, the authors discuss a model of neutrino emission where a fermionic atom turns into a bosonic atom. This possibility has already been ruled out by our paper~\cite{fermion_impossibility}.  Blasone et al. agree that it is ``the change in statistics'' which ``inhibits the cooperative dynamics''. On the other hand, they claim that statistics can be modified e.g. when two fermionic parent atoms form a bosonic condensate of fermion pairs and then present their ``fermionic BEC – bosonic BEC model”. However, it is impossible to modify the statistics for neutrino emission. What matters is the emission process which is described by matrix elements that connect fermionic to bosonic states. Combining two fermions into a bosonic molecule does not change the fundamental nature of neutrino emission. However, in the model presented by Blasone et al., all fermionic jump operators are replaced by bosonic destruction operators for molecules (consisting of two fermions), and this eliminates the fundamental nature of neutrino emission. Their model (eqs. 28 and 29) doesn’t describe neutrino emission.

Our paper~\cite{fermion_impossibility} has rigorously derived that for a system of $N$ particles, the largest possible eigenvalue for the neutrino emission jump rate operator is $\propto N$, ruling out any form of superradiance. This fundamental limit cannot be modified by any form of molecular physics and many-body dynamics as postulated by Blasone et al.

%Even if the molecular model by Blasone et a. would be correct, it would never lead to superradiance. A BEC of $N$ fermion pairs requires the volume of each pair to be smaller than $V/N$ where $V$ is the total volume (the authors even assume the deep BEC regime of much smaller molecules). If neutrino emission creates a free bosons, its overlap with the Bose-Einstein condensate of $N$ bosons has an upper bound of $1/N$, precluding any form of superradiant stimulation. The conclusions of the author that this model (which is wrong by itself, see above) exhibits ``a pronounced superradiant behavior'' is based on an invalid single-mode assumption for the bosonic daughter atoms which incorrectly changes a factor of $1/N$ to unity. \ykl{Do we need this paragraph, since we already showed i nthe previous paragraph that the enhancement is strictly 0?}
 The mistake of Blasone et al. is that they include dynamics within the system after the jump process ( two fermions forming a molecule) into a modified jump operator. This leads to unphysical results. A bosonic jump operator for neutrino emission would occur only in situations like neutrino double decay, but the lifetime for this process is $10^{22}$ years, a trillion times the age of the universe.

In conclusion, neutrino superradiance is practically impossible due to recoil and decay dynamics, and fundamentally impossible since $N$ fermionic amplitudes cannot add up to an intensity $\propto N^2$. Almost all of the $N^2$ interference terms cancel because of the minus sign in fermionic anticommutators~\cite{fermion_impossibility}.

\textit{Acknowledgments}.--- Our research has been supported by NSF (grant No. PHY-2208004), from the Center for Ultracold Atoms (an NSF Physics Frontiers Center, grant No. PHY-2317134), by the Vannevar-Bush Faculty Fellowship (grant no. N00014-23-1-2873), by the Gordon and Betty Moore Foundation GBMF ID \# 12405), and by the Army Research Office (contract No. W911NF2410218).
 \\

\bibliographystyle{apsrev4-2}
\bibliography{mainbib}{}

\end{document}